\documentclass[]{aastex631}
\usepackage{url}

\shortauthors{Namizaki et al.}
\graphicspath{{./}{figures/}}

\begin{document}

\title{A Superflare on YZ Canis Minoris Observed by Seimei Telescope and TESS: Red Asymmetry of H$\alpha$ Emission Associated with White-Light Emission}

\author{Keiichi Namizaki}
\affiliation{Department of Astronomy, Kyoto University, Kitashirakawa-Oiwake-cho, Sakyo-ku, Kyoto 606-8502, Japan; namizaki@kusastro.kyoto-u.ac.jp}

\author[0000-0002-1297-9485]{Kosuke Namekata}
\affiliation{ALMA Project, NAOJ, NINS, Osawa, Mitaka, Tokyo, 181-8588, Japan}

\author[0000-0003-0332-0811]{Hiroyuki Maehara}
\affiliation{Okayama Branch Office, Subaru Telescope, NAOJ, NINS, Kamogata, Asakuchi, Okayama 719-0232, Japan}

\author[0000-0002-0412-0849]{Yuta Notsu}
\affiliation{Laboratory for Atmospheric and Space Physics, University of Colorado Boulder, 3665 Discovery Drive, Boulder, CO 80303, USA}
\affiliation{National Solar Observatory, 3665 Discovery Drive, Boulder, CO 80303, USA}
\affiliation{Department of Earth and Planetary Sciences, Tokyo Institute of Technology, 2-12-1 Ookayama, Meguro-ku, Tokyo 152-8551, Japan}

\author[0000-0001-6653-8741]{Satoshi Honda}
\affiliation{Nishi-Harima Astronomical Observatory, Center for Astronomy, University of Hyogo, Sayo, Hyogo 679-5313, Japan}


\author{Daisaku Nogami}
\affiliation{Department of Astronomy, Kyoto University, Kitashirakawa-Oiwake-cho, Sakyo-ku, Kyoto 606-8502, Japan; namizaki@kusastro.kyoto-u.ac.jp}
\affiliation{Astronomical Observatory, Kyoto University, Sakyo, Kyoto 606-8502, Japan}

\author{Kazunari Shibata}
\affiliation{Kwasan Observatory, Kyoto University, Yamashina, Kyoto 607-8471, Japan}
\affiliation{School of Science and Engineering, Doshisha University, Kyotanabe, Kyoto 610-0321, Japan}








\begin{abstract}

Active M-type stars are known to often produce superflares on the surface. Radiation from stellar (super-)flares is important for the exoplanet habitability, but the mechanisms are not well understood. In this paper, we report simultaneous optical spectroscopic and photometric observations of a stellar superflare on an active M dwarf YZ CMi with the 3.8-m Seimei telescope and the $Transiting\, Exoplanet\, Survey\, Satellite$. The flare bolometric energy was $1.3^{+1.6}_{-0.6} \times 10^{34} \,\rm{erg}$ and H$\alpha$ energy was $3.0^{+0.1}_{-0.1} \times 10^{32} \,\rm{erg}$. The H$\alpha$ emission line profile showed red asymmetry throughout the flare with a duration of $4.6-5.1 \,\rm{hrs}$. 
The velocity of the red asymmetry was $\sim 200-500 \,\rm{km\,s^{-1}}$ and line width of H$\alpha$ was broadened up to $34\pm14\,\rm{\AA}$.
The redshifted velocity and line width of H$\alpha$ line decayed more rapidly than the equivalent width, and their time evolutions are correlated with that of the white-light emission. 
This indicates a possibility that the white light, H$\alpha$ red asymmetry, and H$\alpha$ line broadening originate from nearly the same site, i.e., the dense chromospheric condensation region heated by non-thermal electrons.
On the other hand, the flux ratio of the redshifted excess components to the central components is enhanced one hour after the flare onset.
This may be due to the change of the main source of the red asymmetry to the post-flare loops in the later phase of the flare.

\end{abstract}



\section{INTRODUCTION} \label{sec:intro}

Solar and stellar flares are known to be explosive phenomena that occur on the solar and stellar surface. Flares are observed as rapid increase in a wide range of electromagnetic radiation from gamma rays to radio waves. The mechanism for the solar and stellar flares is thought to be caused by magnetic reconnection in the coronal region \citep[e.g.][]{2002ApJ...577..422S, 2011LRSP....8....6S, 2017ApJ...851...91N}. The magnetic energy stored around spots is converted into thermal, non-thermal and kinetic energy via the magnetic reconnection \citep[e.g.][]{2013PASJ...65...49S, 2013ApJ...771..127N}. In the case of solar flares, the released energy is typically $10^{29}-10^{32}\,\rm{erg}$ \citep[e.g.][]{2011LRSP....8....6S, 2012ApJ...759...71E}. On the other hand, some of stellar flares release more than ten times the energy of the largest solar flares, which are called superflares \citep[e.g.][]{2012Natur.485..478M, 2013ApJS..209....5S, 2014ApJ...792...67C, 2014ApJ...797..121H, 2016ApJ...829...23D, 2019ApJ...876...58N,  2021ApJ...906...72O}. Recently, there is an increasing interest in how these stellar superflares affect habitability of exoplanets, and how and whether the superflares on the Sun can affect the Earth \citep{2013PASJ...65...49S, 2016NatGe...9..452A, 2019ApJ...881..114Y}.

In the standard model of solar flares, the released energy is transported to the chromosphere and photosphere through thermal conduction and accelerated non-thermal energetic particles. The transported energy increases the temperature and gas pressure in the upper chromosphere and ionize the atoms, resulting in the chromospheric emission, such as H$\alpha$ line \citep{2002A&ARv..10..313P}. The heated hot plasma of a temperature of $10^7\,\rm{K}$ flows upward into the coronal magnetic loop, called chromospheric evaporation \citep{1985ApJ...289..414F}. On the other hand, as a back reaction from the upward chromospheric evaporation, downward flows of chromospheric plasmas ($10^4\,\rm{K}$) also happen, called chromospheric condensation. Some aspects of this standard theory can be seen in the following solar H$\alpha$ observations. At the footpoints of the flaring magnetic loops, the H$\alpha$ line is often symmetrically broadened associated with flares \citep[e.g.][]{1962BAICz..13..236S, 1963BAICz..14..234S, 1997ApJS..112..221J}. This broadening is thought to be mainly caused by electric microfield pressure broadening ($Stark$ effect), indicating the enhanced electron density in the chromosphere, or perhaps magnetic turbulent broadening \citep[e.g.][]{2017ApJ...837..125K, 2018A&A...615A..14F}. In addition, the redshifted components in H$\alpha$ line are often observed at the footpoints, called red asymmetry \citep[e.g.][]{1962BAICz..13...37S, 1970SoPh...11..299J, 2003ApJ...596.1347H, 2012PASJ...64...20A}. The generation mechanism of the red asymmetry is thought to be the downward chromospheric condensation \citep[e.g.][]{1984SoPh...93..105I, 1990ApJ...363..318C}. This is usually observed in the impulsive phase of solar flares.
Another cause of redshifted components in H$\alpha$ is post-flare loops where the evaporated plasmas that become cool and dense in the coronal magnetic loops fall into the chromosphere along the magnetic field lines \citep[e.g.][]{1964ApJ...140..746B}. The red asymmetry of H$\alpha$ due to post-flare loops generally occurs in the decay phase of flares since it requires a cooling time due to the thermal radiation and conduction \citep[e.g.][]{2019A&A...624A..96C, 2020A&A...636A.112C}.

Unlike solar observations, stellar flares have been investigated mainly through photometric observations \citep[e.g.][]{2009AJ....138..633K, 2014ApJ...797..121H, 2015ApJ...814...35C, 2016ApJ...829...23D}, and spectroscopic observations are not so many, especially for superflares. 
For example, the difference between the number of flares observed in one paper is more than two orders of magnitude
between photometry (e.g. $>100,000$ in \citealp[][]{2016ApJ...829...23D}) and spectroscopy (e.g. $\sim10-100$ in \citealp[][]{2013ApJS..207...15K, 2020PASJ...72...68N}).
Red asymmetry of chromospheric lines is sometimes observed in the stellar flares.
For example, \citet{1993A&A...274..245H} reported red asymmetries of Balmer lines during a flare of an M dwarf AD Leo, and \citet{2018A&A...615A..14F} found red asymmetries in 32 snapshot frames in flare or quiescent states of 28 M dwarfs.
Furthermore, \citet{2022arXiv220302292W} investigated the time evolution of the red asymmetry and line width of H$\alpha$ line during a flare on an M4-type star and reported that both decay more rapidly than the H$\alpha$ equivalent width. 
They interpreted these features as evidence of flare-driven coronal rain, chromospheric condensation, or a filament or prominence eruption either with a non-radial backward propagation or with strong magnetic suppression.
However, it is not yet known whether the origin of the red asymmetry can be explained by the same model as that of solar flares since stellar flares cannot be spatially resolved and the observed red asymmetry can be superposition of various phenomena.

Another big challenge on solar and stellar flares is an origin of visible continuum emission in flares, called white-light flares. The mechanism has been controversial although the white-light emission is important in solar and stellar-flare physics since the released flare energy is largely distributed to the white-light flare energy \citep{2011A&A...530A..84K}. Based on solar observations, possible emission mechanisms have been proposed as follows: (1) radiation from the chromosphere heated and condensed by non-thermal electrons \citep[e.g.][]{1984SoPh...93..105I}, (2) radiation from the photosphere heated by high-energy protons \citep[e.g.][]{1970SoPh...15..176N}, and (3) re-radiation through radiative backwarming \citep[e.g.][]{1989SoPh..124..303M}. According to solar observations, a similarity between the light curve and spatial location of hard X-rays and white lights suggests that the white-light emission is related to non-thermal electrons \citep[e.g.][]{2006SoPh..234...79H, 2011ApJ...739...96K}. However, the possibilities (1)$\sim$(3) have not been constrained since the observation of the height and spectrum of solar white-light flares is not so easy. 
In the case of stellar white-light flares, broad band spectra are often reported on M dwarfs since the stars themselves are darker than the Sun and the energy scale of flare is very large. 
For example, \citet{1992ApJS...78..565H} reported a continuum spectrum during a superflare on a mid-M dwarf AD Leo with a blackbody spectrum of the temperature of $8500 - 9500\, \rm{K}$. \citet{2010ApJ...714L..98K, 2013ApJS..207...15K, 2016ApJ...820...95K} reported a variety of broad-band white-light spectra on active M dwarfs and suggested that the variety can be explained by radiation from the chromosphere heated and condensed by non-thermal electrons.
\citet{2020PASJ...72...68N} found a correlation between the white-light emission and H$\alpha$ line broadening during a superflare on an AD Leo, indicating a possible connection between the white-light emission and chromospheric condensation.
Simultaneous spectroscopic observations of H$\alpha$ line can be an important tool to unveil the radiation mechanism of the white-light emission of flares.

In this paper, we report spectroscopic and photometric observations of a superflare on an active M dwarf YZ Canis Minoris (YZ CMi) with high precision and high temporal resolution using the Seimei telescope \citep{2020PASJ...72...48K} and $Transiting\, Exoplanet \,Survey \,Satellite$ \citep[TESS; ][]{2015JATIS...1a4003R}. As a result, we found a possible connection between the H$\alpha$ red asymmetry and white-light emission for this superflare, which can provide an opportunity to unveil the mechanism of the red asymmetry and white-light emission at the same time. The data, observations and analysis methods are described in Section \ref{sec:obs}. Results are described in Section \ref{sec:res}, and we discuss the time variation of the line broadening (Section \ref{subsec:broad}) and the red asymmetry in H$\alpha$ line with that of white light (Section \ref{subsec:redasy}). Summary and future works are described in Section \ref{sec:con}.
\begin{figure}[ht!]
\plotone{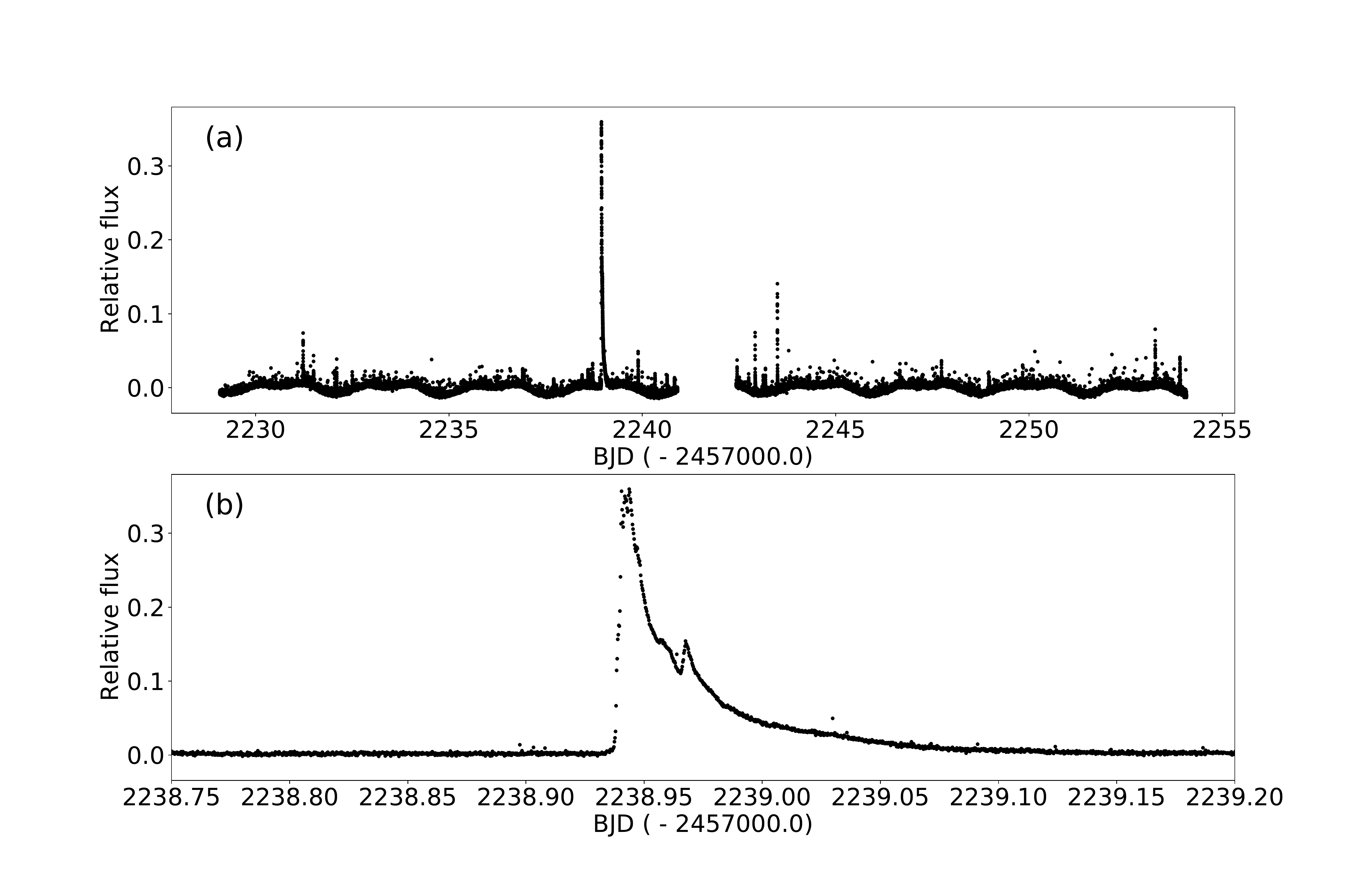}
\caption{Light curve of YZ CMi observed with $TESS$ 20-sec cadence mode. (a) The full light curve during Sector 34. The horizontal and vertical axes represent the observation time in Barycentric Julian Date (BJD) and the relative flux normalized by the average flux ($\delta F/ F_{\rm av}$), respectively. (b) Enlarged light curve around the superflare analyzed in this paper. \label{fig:tess_1}}
\end{figure}


\begin{figure}[ht!]
\plotone{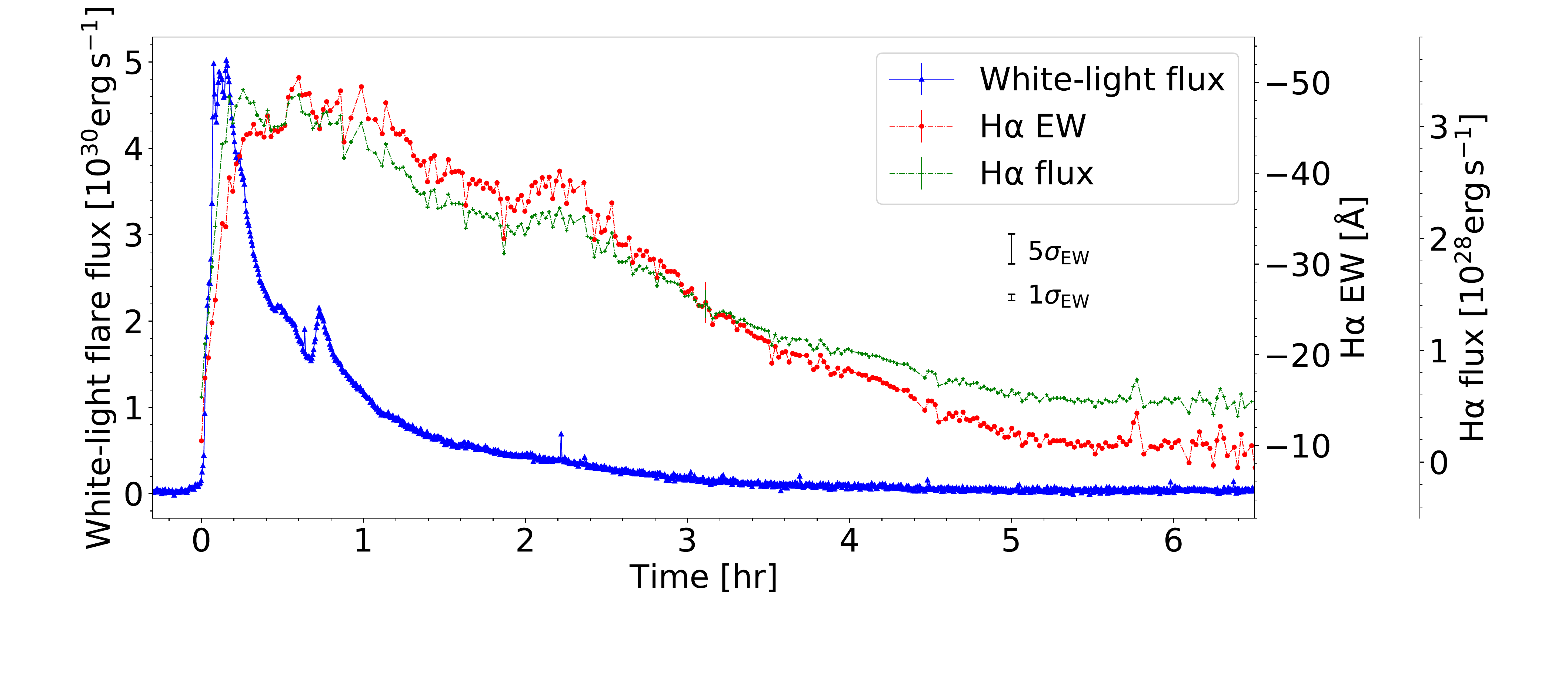}
\caption{Light curves of H$\alpha$ and white-light emission of the superflare on YZ CMi. The blue solid line scaled in the left vertical axis represents the TESS white-light flare flux in the unit of $\rm{erg\,s^{-1}}$. The red and green dashdotted lines represent the H$\alpha$ equivalent width and the H$\alpha$ flux, respectively. The horizontal axis indicates the time since the flare onset (BJD$-2459238.937$) in the unit of hour. The middle and rightmost vertical axes indicate the equivalent width of H$\alpha$ in the unit of $\rm{\AA}$ and the H$\alpha$ flux in the unit of $\rm{erg\,s^{-1}}$. The error bars of the H$\alpha$ equivalent width and H$\alpha$ flux are derived from the residual scattering in the line wing. The $1\sigma_{\rm{EW}}$ and $5\sigma_{\rm{EW}}$ error bar, where $\sigma_{\rm{EW}}$ of 0.7 $\rm{\AA}$ is the standard deviation of the light curve of H$\alpha$ in the quiescent phase (around 6 hrs after the flare onset), are also shown. \label{fig:ew_flux}}
\end{figure}

\begin{figure}[ht!]
\plotone{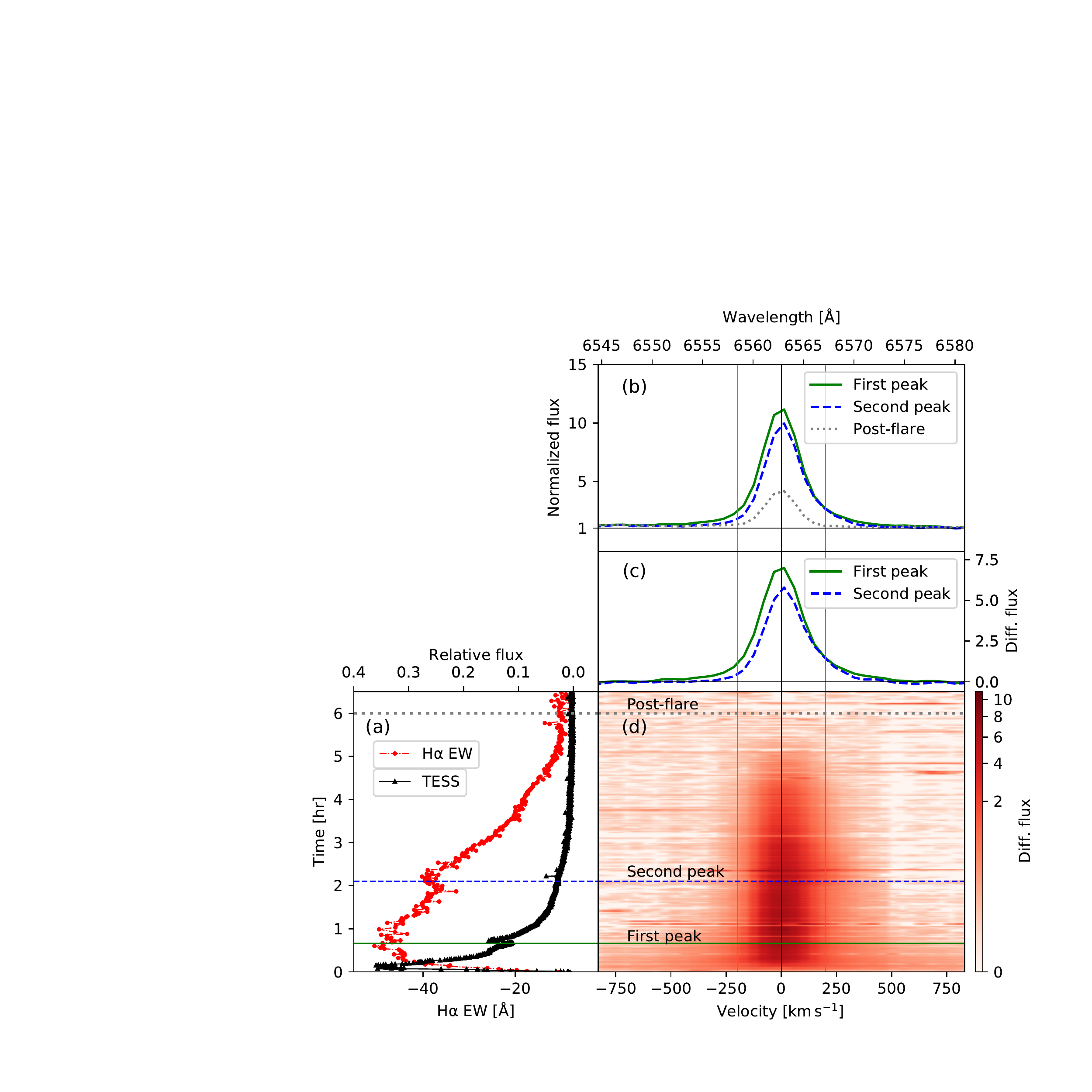}
\caption{Time evolution of the H$\alpha$ spectrum during the superflare on YZ CMi. (a) Red filled circles and black triangles represent the H$\alpha$ equivalent width and white-light emission, respectively. This is basically the same as Figure \ref{fig:ew_flux}. (b) H$\alpha$ line profile. The upper horizontal axis is the wavelength and the vertical axis is the flux normalized by the continuum. Quiescent spectrum of YZ CMi indicated in gray dotted line (the average of 10 spectra around 6 hrs after the flare onset, which correspond to the horizontal grey dotted lines in panels (a, d)). The green solid and blue dashed lines are H$\alpha$ line profiles in $0.67$ hrs, $2.10$ hrs after the flare onset, respectively, which correspond to the horizontal green solid and blue dashed lines in panels (a, d). (c) Flare spectra subtracted from the quiescent spectra (in this paper, we simply call this flare spectra). (d) Time variation of (quiescent-level-subtracted) H$\alpha$ line profile during the flare. The bottom horizontal axis is the $Doppler$ velocity from the H$\alpha$ line center and the vertical axis is the time from the flare onset, which share the left vertical axis of panel (a). The color map represents the differential flux normalized by continuum level as in the color bar. \label{fig:red}}
\end{figure}

\begin{figure}[ht!]
\plotone{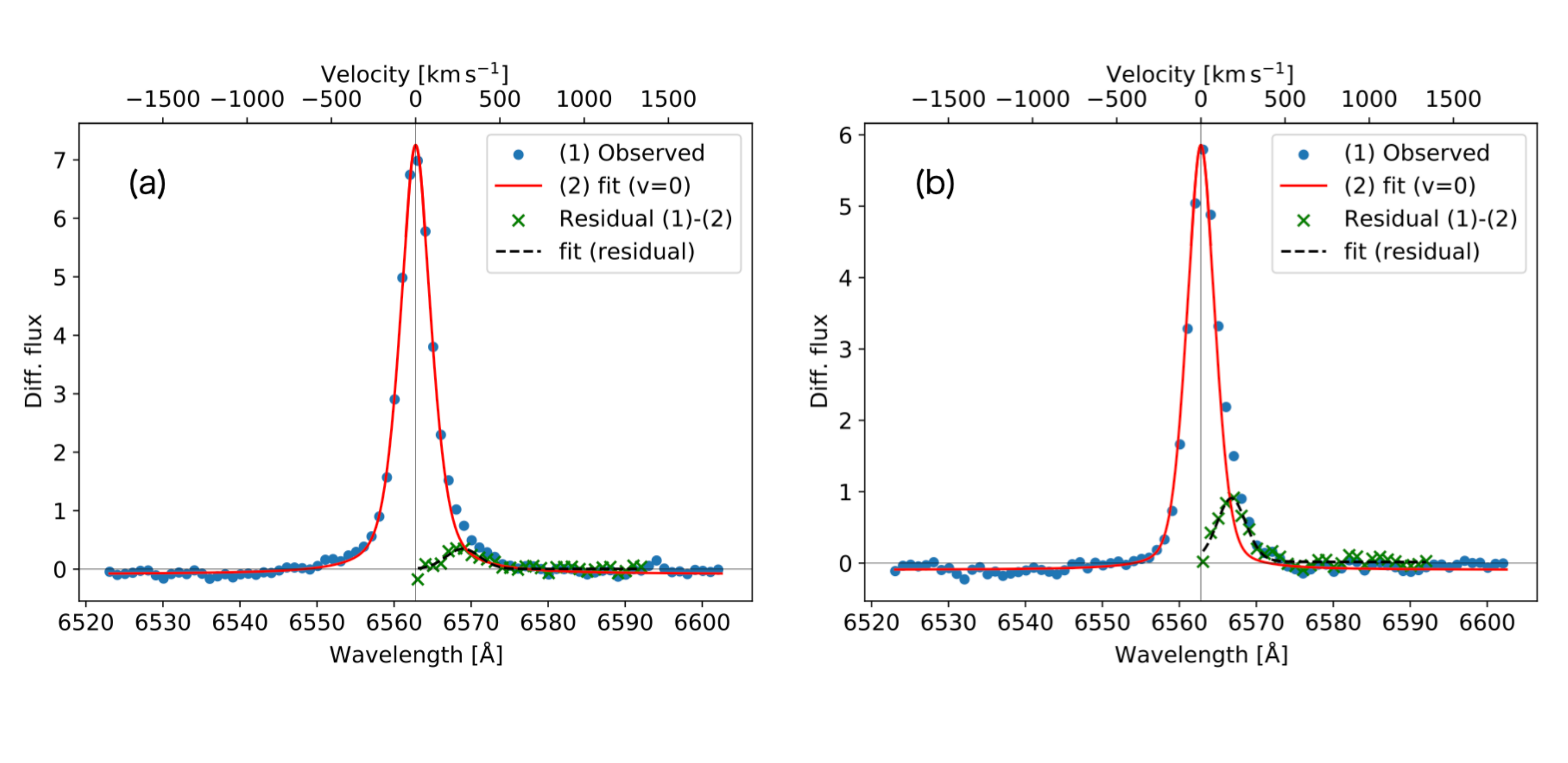}
\caption{Observed H$\alpha$ spectra during the superflare and the fitting results. Panels (a) and (b) represent the spectra indicated with green solid lines ($0.67$ hrs after the flare onset) and blue dashed lines ($2.10$ hrs after the flare onset) in Figure \ref{fig:red}(c). The vertical axis represents the differential flux from the quiescent spectrum. Filled circles represent observed H$\alpha$ line profiles. Red solid lines are symmetric Voigt functions fitted to the components at shorter wavelengths than the H$\alpha$ line center. Green crosses indicate residuals of the central $Voigt$ function from the observed lines. Black dashed lines represent $Voigt$ functions fitted to the residuals (Green crosses). \label{fig:fit}}
\end{figure}

\begin{figure}[ht!]
\plotone{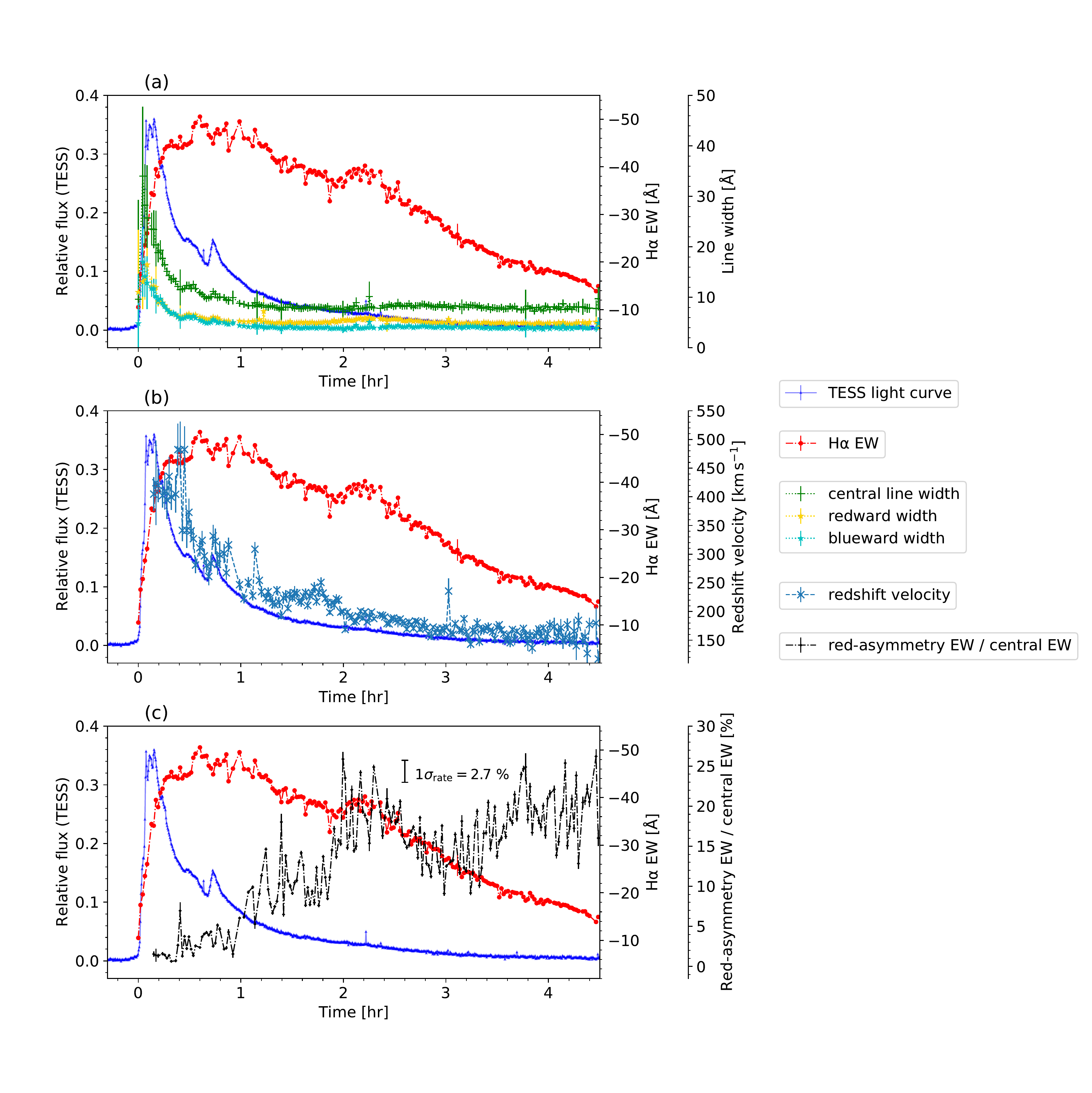}
\caption{H$\alpha$ line width (a), redshift velocity (b), and the equivalent width ratio of the red asymmetry to the central component (c). The blue solid line represents the light curve in white light (cf. Figure \ref{fig:tess_1}), and the red dashdotted line represents the H$\alpha$ equivalent width. The horizontal axis represents the time in unit of $\rm{hr}$. The left and middle vertical axes represent the relative flux normalized by the average flux and the equivalent width in unit of $\rm{\AA}$. They are common in the panels (a), (b) and (c). (a) The green dotted line with cross symbols scaled in the rightmost vertical axis indicate central line width of H$\alpha$ in unit of $\rm{\AA}$. Yellow and cyan dotted lines with star symbols represent line width of red and blue side, respectively (i.e., $2\times$cyan = green). The error bars are calculated from the fitting errors as the uncertainties. (b) The darkcyan dashed line with cross symbols scaled in the rightmost vertical axis indicate the line-of-sight velocity of redshifted excess components (derived from the central wavelength of black dashed line in Figure \ref{fig:fit}) with the error bars calculated from the fitting errors. (c) The black dashdotted line with diamond symbols scaled in the rightmost vertical axis indicate the ratio of the equivalent width of redshifted excess components to the H$\alpha$ equivalent width with the error bars calculated from the fitting errors. The $1\sigma_{\rm{rate}}$ error bar is the standard deviation of the residuals calculated from the moving average. The data after 4.5 hrs are not plotted, since the fitting could not be done well because of the weak H$\alpha$ emission. In addition, the values of redshift velocity (b) and the ratio of the equivalent width (c) within 9 minutes from the flare onset were also excluded because the values obtained by the second fitting to redshifted components were not reliable due to the large variances of the residuals. \label{fig:s-t}}
\end{figure}

\section{OBSERVATIONS \& ANALYSES} \label{sec:obs}

\subsection{Target star : YZ CMi \label{subsec:obs_y}}

The active M4.5 dwarf YZ CMi has a thick convection zone and a rapid rotation with a period of 2.8 days, which results in a high flare frequency \citep[e.g.][]{2008MNRAS.390..567M}. Numerous flares on YZ CMi have been observed in a wide range of wavelengths from radio to X-ray since the first flare in this object was observed by \citet{1945PASP...57..216V}.  In particular, a superflare with an U-band energy of $\sim10^{34} \,\rm{erg}$ was reported in \citet{2010ApJ...714L..98K}. \citet{1976ApJS...30...85L} and \citet{2021PASJ...73...44M} measured a flare-frequency distribution of YZ CMi and found that its power-law index is comparable to those of other M-type stars and the Sun.

\subsection{TESS \label{subsec:obs_t}}

$TESS$ performed optical photometric observations of YZ CMi in the wavelength of $6,000-10,000\,\rm{\AA}$, simultaneously with our ground-based spectroscopic observations. $TESS$ was launched in April 2018 and has four cameras covering $24^\circ \times 96^\circ$ degree strips of the sky called Sectors. Each Sector observes the area for about $27$ days on average and the data obtained by $TESS$ are of very high-precision. $TESS$ observation for Sector 34 was performed from January 14, 2021 to February 8, 2021 (Barycentric Julian Date (BJD) : $2459229.09-2459254.07$). In Sector 34, $TESS$ observed YZ CMi with 20-sec cadence. We analyzed the $TESS$ Pre-search Data Conditioned Simple Aperture Photometry (PDC-SAP) light curves retrieved from the MAST Portal site  (\url{https://mast.stsci.edu/portal/Mashup/Clients/Mast/Portal.html}). The normalized $TESS$ light curve shown in Figure \ref{fig:tess_1}(a) shows rotational brightness variations and rapid flares, and Figure \ref{fig:tess_1}(b) is an expanded panel of a superflare of interest in this paper. 

From the $TESS$ light curve, we estimated the energy of the white-light flare by assuming $10,000\,\rm{K}$ blackbody radiation \citep{2013ApJS..209....5S}. First of all, the flare flux ($\delta F$) normalized by the averaged stellar flux ($F_{\rm{av}}$) can be expressed as the ratio of the flare luminosity to stellar luminosity as follows: 
\begin{eqnarray}
\frac{\delta F}{F_{\rm{av}}} = \frac{A_{\rm{flare}}}{\pi R_{\rm{star}}^2}\frac{\int R_{\lambda} B_{\lambda}\left(T_{\rm flare}\right)d\lambda}{\int R_{\lambda} B_{\lambda}\left(T_{\rm eff}\right)d\lambda},
\label{eq:1}
\end{eqnarray}
where $\delta F/F_{\rm av}$ is the relative flux, $R_{\lambda}$ is the response function of the $TESS$ detector, $R_{\rm star}$ is the radius of YZ CMi ($\sim 0.3\, R_{\rm solar}$), $A_{\rm flare}$ is the flaring area, $T_{\rm eff}$ is the stellar effective temperature of $\sim3,300\,\rm{K}$ \citep{2014ApJ...791...54G, 2015ApJ...800...85N, 2016ApJ...822...97H}, $T_{\rm flare}$ is the flare temperature of $10,000\,\pm\,5,000 \,\rm{K}$ \citep[e.g.][]{1992ApJS...78..565H, 2020ApJ...902..115H}, and $B_{\lambda}$ is the Planck function. After deriving the flaring area based on the Equation \ref{eq:1}, we calculated the flare luminosity using the following equation,
\begin{eqnarray}
L_{\rm flare} = \sigma T_{\rm flare}^4 A_{\rm flare},
\end{eqnarray}
where $\sigma$ is the Stefan-Boltzmann constant.
By integrating the flare luminosity along time, the bolometric energy of the superflare ($E_{\rm bol}$) was obtained. Since $E_{\rm bol}$ is mainly dependent on $T_{\rm flare}$, the uncertainty of $E_{\rm bol}$ was estimated mainly by considering the error of $\pm\,5,000 \,\rm{K}$ \citep[][]{2020ApJ...902..115H}.
In addition, the duration of the white-light flare was calculated as the decay time that is at one-tenth of the peak in the light curve.

\subsection{3.8-m Seimei telescope \label{subsec:obs_s}}

We performed spectroscopic observations of YZ CMi with the 3.8-m Seimel telescope located at Okayama Observatory, Japan. 
Seimei telescope is equipped with a low-dispersion spectrograph, KOOLS-IFU \citep[Kyoto Okayama Optical Low dispersion Spectrograph with an optical-fiber Integral Field Unit;][]{2019PASJ...71..102M}. We used the VPH683 grism covering the wavelength range of $5800-8000 \,\rm{\AA}$ with a spectral resolution of $\lambda/{\Delta \lambda}\sim2000$. This wavelength coverage includes the H$\alpha$ line ($6562.8\, \rm{\AA}$), which is a chromospheric line often used as a tracer of flaring activity. The exposure time was $60$ sec and the CCD readout time was about 17 sec. The spectroscopic observation with this high temporal resolution allows us to investigate the temporal variation of the H$\alpha$ spectrum during stellar flares in detail. 

With the aim of investigating the correspondence between the white light and H$\alpha$ in stellar flares, we performed simultaneous observations of YZ CMi in accordance with $TESS$ sector 34. 
The data reduction follows the prescription in \citet{2020PASJ...72...68N, 2022ApJ...926L...5N, 2022NatAs...6..241N} with IRAF and PyRAF packages, and each spectrum was calibrated by considering the radial velocity of YZ CMi ($26.495\,\rm{km\,s^{-1}}$; \citealp[][]{2018A&A...616A...7S}).
From the H$\alpha$ line profile, we measure the H$\alpha$ equivalent width (EW) which is H$\alpha$ emission integrated for $6562.8\pm 15\,\rm {\AA}$ after being normalized by the nearby continuum level, and then plotted the light curve in Figure \ref{fig:ew_flux}. In addition, the H$\alpha$ flux at each time was calculated by simply multiplying the equivalent width by the continuum flux near the H$\alpha$. 
The continuum flux was obtained based on a flux calibrated quiescent spectrum \citep{2013ApJS..207...15K} and the same parameters used when deriving bolometric energy. 
Then, the H$\alpha$ energy of the flare ($E_{\rm H\alpha}$) and its uncertainty were derived by integrating it along the time considering the error of $T_{\rm flare}$.
Furthermore, the duration of the H$\alpha$ flare was calculated as the decay time for the H$\alpha$ EW to decay to the quiescent level with 1 or 5 sigma, which is the standard deviation of the residuals.

\subsection{Analysis of the red asymmetry and line broadening of the H$\alpha$ line profile \label{subsec:obs_red}}

We here introduce the analysis of fitting of the flare spectrum in order to characterize the line broadening and asymmetry of H$\alpha$ line in stellar flares. The example of flare spectra are shown in Figure \ref{fig:red}. We normalized the H$\alpha$ line profile by the continuum level and subtracted the quiescence spectrum (gray dotted line in Figure \ref{fig:red}(b)) from the flare spectra (e.g., green solid and blue dashed line in Figure \ref{fig:red}(b)). Here, we take the spectrum as the quiescence spectrum at around $6$ hour in Figure \ref{fig:ew_flux} since the equivalent width at this time and that on other days when flares did not occurred are at the nearly same level. We then obtained the time variation of the quiescence-level-subtracted H$\alpha$ line profile as in Figure \ref{fig:red}(c, d). 

We derived the line width and velocity of the asymmetric components. In the following, we introduce the fitting method when the flare spectrum shows a redshifted component in the H$\alpha$ wing. First, we fitted only the blueward line profile from the H$\alpha$ line center with a $Voigt$ function (see, red solid line in Figure \ref{fig:fit}), with the aim of simply estimating the line profile without a redshift component. The function obtained by fitting the blueward line profile was folded back to the red side at the H$\alpha$ line center (see, red solid line in Figure \ref{fig:fit}). We then subtracted the obtained symmetrically broadened line (i.e. the model of a non-moving spectrum with a velocity of 0) from the observed line profile (blue points in Figure \ref{fig:fit}). The resulting residuals (green crosses in Figure \ref{fig:fit}) in the redward asymmetry is regarded as the ``redshifted component" in this study. We again fitted a $Voigt$ function to the residuals and estimated the redshifted component (black dashed lines in Figure \ref{fig:fit}) with a free parameter of the central velocity. 
Here, if the central components were fitted simultaneously with redshifted components, the fitting did not work well and the redshifted components were not extracted properly. Thus, we decided to fix the central component before fitting the redshifted components.
As a result of this fitting, the line-of-sight redshifted velocity is estimated from the $Doppler$ shift of the redward residual component (black dashed lines in Figure \ref{fig:fit}). The equivalent width is also calculated for the central components and redshifted components by simply integrating along wavelengths, and their ratio is also obtained. This method was applied to each spectrum observed with the time interval of $\sim77\,\rm{sec}$. In the same way, spectra showing ``blue asymmetry" can be fitted, but we could not see the ``blue asymmetry" in this flare as will be discussed in Section \ref{sec:res}. Only the redward excess was then evaluated in this study. 

In addition, with the aim of estimating the blueward and redward line broadening, we performed $Voigt$ function fittings for the blueward and redward line profiles separately. The blueward and redward line widths were calculated by obtaining the wavelengths at one-eighth of the peak intensity for the blue and red sides, respectively. In this paper, we define the central line width (The green dotted line with cross symbols in Figure \ref{fig:s-t}(a)) as twice the blueward line width (cyan dotted lines with star symbols in Figure \ref{fig:s-t}(a)).

Futhermore, the decay times of the central line width and redshift velocity were calculated at the time when they decayed to one-tenth of the peak. Here, both standard deviations of the residuals ($\sigma_{\rm{width}}$, $\sigma_{\rm{velocity}}$) were added to the one-tenth of the peak values as errors to estimate uncertainties of each decay times.

Note that the line width and redshift velocity near the spectral resolution limit of KOOLS-IFU ($\sim 150 \,\rm{km\,s^{-1}}$) could be overestimated, although the qualitative trend would be real. For this reason, we do not discuss quantitatively the data after $3\,\rm{h}$ in Figure \ref{fig:s-t}(c) in this paper.
\section{RESULTS AND DISUCUSSION} \label{sec:dis}
\subsection{Summary} \label{sec:res}

Figure \ref{fig:tess_1} shows the $TESS$ light curve of YZ CMi in Sector 34. The flare of interest in this paper is the largest event among those detected by $TESS$ Sector 34 observations. As shown in Figure \ref{fig:tess_1}(b), the white-light increases by $\sim35\,\%$ relative to the averaged flux of the quiescence phase. We successfully observed this largest event in $TESS$ Sector 34 by ground-based optical spectroscopic observations. The bolometric energy of the flare ($E_{\rm bol}$) was calculated as $1.3^{+1.6}_{-0.6}\times10^{34} \,\rm{erg}$ by using the method described in Section \ref{subsec:obs_t}; this can be classified to be a superflare. 
Here, it was difficult to constrain $T_{\rm{flare}}$ by fitting to spectrum because of the narrow wavelength range and lack of NUV spectrum covered by VPH683 grism. However, even assuming a $20,000 \,\rm{K}$ blackbody, $E_{\rm bol}$ is $5.5\times10^{34} \,\rm{erg}$, and thus it is not significantly dependent on $T_{\rm{flare}}$.
Figure \ref{fig:ew_flux} shows the light curves of the superflare in $TESS$ white light and H$\alpha$. It is found that the white-light flare is dominant in the rising phase of the H$\alpha$ equivalent width and decay more rapidly (in $1.6\,\rm{hr}$) than the H$\alpha$ flare (in $4.0-4.5\,\rm{hr}$). 
It is known that white light is sensitive to the non-thermal heating and the H$\alpha$ radiation is related to both non-thermal and thermal heating \citep[e.g.][]{2020PASJ...72...68N}. Therefore there is a possibility that the observed time decay between H$\alpha$ and white light could be the indication of the $Neupert$ effect in solar flares, which is often seen in soft and hard X-ray radiation \citep{1968ApJ...153L..59N, 1993SoPh..146..177D}.
The H$\alpha$ energy of the flare ($E_{H_{\alpha}}$) was derived as $3.0^{+0.1}_{-0.1}\times10^{32}\,\rm{erg}$, which is $\sim 2.3 \,\%$ of the bolometric energy of this superflare. 

Figure \ref{fig:red} shows the temporal evolution of the H$\alpha$ spectrum. The H$\alpha$ line profile is widely broadened in the impulsive phase (hereafter the ``impulsive phase" means the time between 0 hr and $\sim 1\,\rm{hr}$ from the flare onset), while it narrows in the decay phase. In addition, red asymmetries can be seen throughout the flare (e.g. Figure \ref{fig:red}(c, d)). 
The results of spectral fitting for the blue solid and green dashed lines in Figure \ref{fig:red} are shown in Figure \ref{fig:fit}(a) and (b), respectively. For example, the central line width and redshift velocity of Figure \ref{fig:fit}(a) are estimated as $9.7\pm0.5 \,\rm{\AA}$ and $280\,\pm\,12\,\rm{km\,s^{-1}}$, respectively, and those of Figure \ref{fig:fit}(b) are $8.3\pm0.4\,\rm{\AA}$ and $181\pm5\,\rm{km\,s^{-1}}$, respectively. 

Figure \ref{fig:s-t} shows the temporal evolution of the fitting parameters of (a) the line width, (b) redshift velocity, and (c) equivalent width ratio of the red asymmetry to the central component, in comparison with the white-light and H$\alpha$ flare light curves (blue and red solid lines). The features are summarized as follows:
\begin{enumerate}
\item[(a)] The central line width of H$\alpha$ line peaks at $34\pm14\,\rm{\AA}$ ($0.04-0.12$ hrs after the flare onset) and decays rapidly to $10.6\pm0.4\,\rm{\AA}$ (one-tenth of the peak value) in $0.6-0.7\,\rm{hr}$. We found that the temporal evolution of the central line width is similar to the white-light flare, including the small brightness variation, and decays more rapidly than the H$\alpha$ equivalent width. Both blueward and redward wings show the same trend.
\item[(b)] The redshift velocity peaks at $482\pm40\,\rm{km\,s^{-1}}$ ($0.39-0.45$ hrs after the flare onset) and decays rapidly to $183\pm9\,\rm{km\,s^{-1}}$ (one-tenth of the peak value) in $1.6-2.3\,\rm{hr}$. The temporal evolution of the redshift velocity is also similar to the white-light flare, and decays more rapidly than the H$\alpha$ equivalent width. This is very similar to the feature of the central line width (a).
\item[(c)] The equivalent width ratio of the red asymmetry to the central component is small in the impulsive phase (a few per cent) and increases up to $\sim 20\,\%$ in the decay phase.
\end{enumerate}
 


\subsection{Relation between line broadening and white-light emission \label{subsec:broad}}

During the impulsive phase ($0-1\,\rm{hr}$ in Figure \ref{fig:s-t}(a)), H$\alpha$ line is almost symmetrically broadened, associated with white-light emission. There could be a possible redshifted excess in this impulsive phase, but in the first 20 min, the redshifted excess is too weak to be fitted, and therefore we call the profiles in this phase almost symmetric. This means that line broadening of H$\alpha$ and white-light emission are related to the same region. Line broadening of H$\alpha$ is often thought to be caused by $Stark$ effect, indicating that the electron density of the chromosphere becomes very high. \citet{2020PASJ...72...68N} suggested that line broadening of H$\alpha$ flare with a width of more than $10 \,\rm{\AA}$ cannot be explained only by thermal conduction heating, but by non-thermal electron heating of the deep chromosphere. Thus, the line broadening observed in our paper would indicate that the deep chromosphere is heated by non-thermal electrons during this superflare.
Furthermore, the time evolution of the central line width of H$\alpha$ and that of white light show a very good correlation, which suggests that the white-light emission is also radiated from the non-thermally heated chromosphere in this superflare.
\citet{2015SoPh..290.3487K} simulated and discussed a feasibility that white-light flares are radiated from the chromospheric condensation region in the case of an M-dwarf atmosphere. They found that a photosphere-like dense region is generated in the chromosphere heated by non-thermal electrons, producing significant white-light emission.  As described in the above scenario, 
our observations suggest a possibility that the white-light emission and line broadening are produced in the chromosphere heated by non-thermal electrons during the impulsive phase.
Alternatively, it is possible that not only non-thermal electrons but also non-thermal protons accelerated by flares may produce a similar behavior, although their properties are not well studied both observationally and theoretically. On the other hand, the radiative backwarming scenario (proposed by \citealp[][]{1989SoPh..124..303M}) would be less likely in this case because if the backwarming is dominant (and non-thermal heating is less dominant), the time evolution of the white-light flare would be similar to that of the H$\alpha$ flare because both are heated by the same thermal source in the coronae, i.e., the soft X-ray loops (proposed by \citealp[][]{2022ApJ...926L...5N}), which is different from our observation. The possibility of the radiative backwarming can be further constrained with future simultaneous observations with soft X-ray, which is the main source of irradiation from the coronae \citep{2019MNRAS.489.4338N}.

Let us discuss quantitative parameters of this superflare. \citet{2020PASJ...72...68N} reported a $2\times 10^{33}\,\rm{erg}$-class superflare with a strong H$\alpha$ line broadening on an M-type flare star AD Leo. In the event, the H$\alpha$ emission line full width at 1/8 maximum increased up to $14 \,\rm{\AA}$ in the initial phase and showed a slow decay compared to the spiky increase. This feature is similar to that of our superflare although the energy scale is a bit different ($1\times 10^{34}\,\rm{erg}$ in our case). \citet{2020PASJ...72...68N} suggested that the line width can be explained by a non-thermal electron beam of $\sim 10^{12}\, \rm{erg\,cm^{-2}\,s^{-1}}$ injected into the chromosphere. 
On the other hand, our superflare temporarily showed the full width at 1/8 maximum of $34\pm14 \,\rm{\AA}$, suggesting an injection of a much higher flux of a non-thermal electron beam. In addition, the central line width dramatically decreases from $34\pm14 \,\rm{\AA}$ in the initial phase to one-tenth of its peak value in the decay phase within one hour. This means that the non-thermal heating also dramatically decreases as the flare decays. Considering that the variation timescale is very long ($0.6-0.7 \,\rm{hr}$) compared to the flare heating timescale at a given flare ribbon seen in solar flares, which is a few ten seconds to 10 minutes at the most \citep{2006ApJ...641.1210X, 2016ApJ...833...50K}, we expect that this does not represent the decreasing flare heating at a given flare kernel but would represent the decreasing flare heating at flare ribbons moving outward as a result of successive reconnection.
\subsection{Interpretation of red asymmetry \label{subsec:redasy}}

In this section, we discuss the origin of the red asymmetry during our superflare. As shown in Section \ref{sec:res}, red asymmetry with a large velocity of $482\pm40 \,\rm{km\,s^{-1}}$ was observed during the impulsive phase and decayed more rapidly in $1.6-2.3$ hr than the H$\alpha$ equivalent width (in $4.0-4.5$ hr). The velocity of the redshifted excess components slows down on a timescale similar to those of the white-light emission and line broadening (Figure \ref{fig:s-t}(b)). As discussed in Section \ref{subsec:broad}, evolution of both the white-light emission and line broadening could be related to that of the chromospheric condensation region. Hence, this similarity indicates that the chromospheric condensation can be the origin of the red asymmetry, especially in the impulsive phase. In the case of solar flares, it is often observed that velocities of redshifted components peak out and decay more rapidly than H$\alpha$ intensities in the footpoints (i.e. chromospheric condensation regions) \citep{1984SoPh...93..105I}. This is considered to indicate the properties of the chromospheric condensation process, where thickness of the chromospheric condensation increases with the rapid decrease of its velocity \citep{1981SoPh...73..269L, 1982SoPh...81..281S, 1984SoPh...93..105I, 2022arXiv220113349K}. Likewise, in our superflare, the velocity of the redshifted components also decayed more promptly than the H$\alpha$ equivalent width, which supports the possibility that the chromospheric condensation can be the origin of the red asymmetry. 

The initial velocity of the redshifted components $482\pm40 \,\rm{km\,s^{-1}}$ is faster than those of red asymmetry of solar flares ($\leq 100\,\rm{km\,s^{-1}}$) \citep[e.g.][]{1984SoPh...93..105I, 2012PASJ...64...20A}. 
In the case of stellar flares, the typical redshift velocity is $100-200\,\rm{km\,s^{-1}}$ \citep[e.g.][]{1993A&A...274..245H, 2022arXiv220302292W}.
\citet{2021A&A...646A..34K} analyzed the asymmetries of H$\alpha$ lines during the flares on M-type stars and detected the maximum velocities of $358\,\rm{km\,s^{-1}}$.
Compared to these studies, the observed initial velocity of redshifted components is relatively fast.
In numerical simulations assuming high electron-beam fluxes associated with active M-dwarf flares, most studies show that the downward velocity of the chromospheric condensation is at most $\sim 100\,\rm{km\,s^{-1}}$ \citep{2006ApJ...644..484A, 2017ApJ...836...12K, 2022ApJ...928..190K}.
\citet{2014ApJ...795...10L} investigated the atmospheric response to extremely high-flux beams by numerical simulations and suggested a possibility of the fast condensation velocity of $100-1,000\,\rm{km\,s^{-1}}$. 
Therefore, the observed redshift velocity of $200-500\,\rm{km\,s^{-1}}$ in our superflare can be still explained by the chromospheric condensation model with large energy inputs.
However, note that the observed fast velocities could be possibly due to the two-components fitting method employed in this study since it is pointed out that this method tends to overestimate the velocity \citep{2022ApJ...933..209N}. 
For example, the peak velocity of $\sim482\,\rm{km\,s^{-1}}$ is expected to be roughly overestimated by the redward line width of $6.2\,\rm{\AA}$ ($288\,\rm{km\,s^{-1}}$). 
Considering this overestimation, the actual maximum redshift velocity may be $\sim200\,\rm{km\,s^{-1}}$ at most.
In any case, the uncertainty does not affect the qualitative discussion since the trend of time variation does not change, even if the velocity of the redshifted components is overestimated.

After the impulsive phase, the red asymmetry in H$\alpha$ still lasted until $4.6-5.1$ hr from the flare onset with a velocity of $\sim 150-300 \,\rm{km\,s^{-1}}$ (Figure \ref{fig:red}(d) and \ref{fig:s-t}). In addition, Figure \ref{fig:s-t}(c) shows that the equivalent width ratio of the red asymmetry to the central component significantly increased in 1-3 hrs from the flare onset
after the equivalent width of the central component decayed. If the main source of the red asymmetry in the decay phase ($> 1\,\rm{hr}$) is also the additionally heated chromospheric condensation region, we would expect to see the increase of blueward line broadening and white-light enhancement. However, as shown in Figure \ref{fig:s-t}(a), there was no additional increase in blueward line broadening and white-light emission during the decay phase, while only the redward line broadening increased. Therefore, it is possible that a physical mechanism other than chromospheric condensation is at work, especially in the decay phase. We suggest that the red asymmetry seen in the late phase of our superflare might be explained by post-flare loops. \citet{1998A&A...337..911H} reported that some solar flares maintained post-flare loops in H$\alpha$ for more than several hours. Furthermore, recent simulations suggest that it takes $\sim30-40\,\rm{min}$ that typical flare loops cool down to $\sim10^4\,\rm{K}$ \citep{2013ApJ...772...40C, 2020ApJ...890..100R, 2021ApJ...920L..15R}. Based on these solar observations, the observed red asymmetry of this superflare that lasted for several hours and delayed $\sim 1$ hr compared to impulsive phase can be representative of the nature of post-flare loops.
\section{SUMMARY \& FUTURE WORKS} \label{sec:con}

We performed simultaneous optical spectroscopic and photometric observations of an active M-type flare star YZ CMi and successfully detected a superflare with the energy of $1.3^{+1.6}_{-0.6} \times 10^{34} \,\rm{erg}$ and duration of $4.6-5.1$ hrs. Since the number of spectroscopic observation of superflares is still small, our observation with high time cadence and high precision is precious to reveal the superflare mechanism. 

During the initial phase, significant line broadenings of H$\alpha$ line with widths of up to $34\pm14\,\rm{\AA}$ were observed, indicating the dense chromospheric region heated by high-flux non-thermal electron beams. The temporal evolution of the line broadening showed a good correspondence with that of the white-light flare, which means that the white-light emission is also radiated from the non-thermally heated chromosphere in this superflare. In addition, red asymmetry was observed in H$\alpha$ line almost throughout the flare. The velocity of the redshifted excess components with the initial velocity of $482\pm40\,\rm{km\,s^{-1}}$ slows down on a timescale similar to those of the white-light emission and line broadening, and decayed more rapidly than H$\alpha$ equivalent width. These indicate a possibility that the chromospheric condensation can be the origin of the observed white-light emissions and red asymmetry in the impulsive phase.
On the other hand, during the late phase, the post-flare loops can also contribute to the red asymmetry since the equivalent width ratio of the red asymmetry to the central component increased as the flare decayed.
Our simultaneous spectroscopic and photometric observations provide further constraints on the mechanism responsible for the red asymmetry during (super-)flares. 

Moreover, the light curves of the white light and H$\alpha$ show several secondary peaks during the flare (e.g. Figure \ref{fig:ew_flux}(a)). Interestingly, the white-light emission of this superflare showed quasi-periodic pulsations \citep[QPP; e.g.][]{1983ApJ...271..376K} in its light curve, although we did not analyze them in this paper. In future work, we will compare the quasi-periodic pulsations of white-light flares and time variation of H$\alpha$ line. In addition, since most of the previous solar observations referred in this paper focus on spatially resolved information on a small portion of the solar disk, we do not in detail know how the H$\alpha$ line profile behaves when solar flares are spatially integrated on the Sun seen as a star \citep[cf.][]{2022ApJ...933..209N, 2022ApJ...939...98O}. Therefore, in the future, we will conduct Sun-as-a-star analyses of not only flare kernels but also post-flare loops to improve the interpretation of stellar data in this paper. 

Finally, we briefly comment on a possible relationship between our superflares and stellar coronal mass ejections (CMEs). CMEs are phenomena that carry plasmas into interplanetary space by ejecting plasmas upward through the release of magnetic energy, and are thought to have a significant impact on the planetary habitability and stellar evolution \citep[e.g.][]{2011SoPh..268..195A, 2013ApJ...764..170D, 2020IJAsB..19..136A}. Some study reported that stellar flares sometimes show blueshifted emission or absorption profiles in the H$\alpha$ line \citep[e.g.][]{2019A&A...623A..49V, 2021PASJ...73...44M, 2022NatAs...6..241N}. These are often interpreted as stellar prominence eruptions, which are indirect evidence of the CMEs.
In our superflare, only strong redshifted components were observed, and significant blueshifted components were not observed at all times during the flare. In addition, since a white-light flare was observed, it is speculated that the flare footpoints was on the visible side of the stellar hemisphere. Thus, a possibility of a backward-directed CME related to a flare occurring close to the limb is unlikely for our superflare. There can be a possibility that strong overlying magnetic fields of the M dwarf may have suppressed a prominence eruption and CME, resulting in a significant reduction of velocity \citep{2018ApJ...862...93A}. Moreover, it is possible that the strong chromospheric condensation during the impulsive phase of this superflare could have masked the faint blueshifted components associated with weak CMEs \citep{2021A&A...646A..34K, 2022ApJ...939...98O}. We expect that a larger sample of time-resolved spectroscopic observations, such as those in this study, and simultaneous observations of X-ray and ultraviolet band will reveal more details about the relationship between redshifts (and blueshifts) and CMEs on stars in the future.






\begin{acknowledgments}
This study makes use of data obtained through the programs, 21A-N-CN03 and 21A-K-0009 in the open-use of the observing time at the 3.8-m Seimei telescope provided by NAOJ. Funding for
the TESS mission is provided by NASA’s Science Mission directorate. We acknowledge the International Space Science Institute and the supported International Team 464: The Role of Solar and Stellar Energetic Particles on (Exo) Planetary Habitability (ETER- NAL, \url{http://www.issibern.ch/teams/exoeternal/}). Y.N. was supported by the JSPS Overseas Research Fellowship Program. This research is supported by JSPS KAKENHI grant numbers 18J20048, 21J00316 (K.N.), 21J00106 (Y.N.), and 21H01131 (M.H., D.N., K.S.). Y.N. was also supported by NASA ADAP award program Number 80NSSC21K0632 (PI: Adam Kowalski)
\end{acknowledgments}

\bibliography{ms}{}

\begin{thebibliography}{}
\expandafter\ifx\csname natexlab\endcsname\relax\def\natexlab#1{#1}\fi
\providecommand{\url}[1]{\href{#1}{#1}}
\providecommand{\dodoi}[1]{doi:~\href{http://doi.org/#1}{\nolinkurl{#1}}}
\providecommand{\doeprint}[1]{\href{http://ascl.net/#1}{\nolinkurl{http://ascl.net/#1}}}
\providecommand{\doarXiv}[1]{\href{https://arxiv.org/abs/#1}{\nolinkurl{https://arxiv.org/abs/#1}}}

\bibitem[{{Aarnio} {et~al.}(2011){Aarnio}, {Stassun}, {Hughes}, \&
  {McGregor}}]{2011SoPh..268..195A}
{Aarnio}, A.~N., {Stassun}, K.~G., {Hughes}, W.~J., \& {McGregor}, S.~L. 2011,
  \solphys, 268, 195, \dodoi{10.1007/s11207-010-9672-7}

\bibitem[{{Airapetian} {et~al.}(2016){Airapetian}, {Glocer}, {Gronoff},
  {H{\'e}brard}, \& {Danchi}}]{2016NatGe...9..452A}
{Airapetian}, V.~S., {Glocer}, A., {Gronoff}, G., {H{\'e}brard}, E., \&
  {Danchi}, W. 2016, Nature Geoscience, 9, 452, \dodoi{10.1038/ngeo2719}

\bibitem[{{Airapetian} {et~al.}(2020){Airapetian}, {Barnes}, {Cohen},
  {Collinson}, {Danchi}, {Dong}, {Del Genio}, {France}, {Garcia-Sage},
  {Glocer}, {Gopalswamy}, {Grenfell}, {Gronoff}, {G{\"u}del}, {Herbst},
  {Henning}, {Jackman}, {Jin}, {Johnstone}, {Kaltenegger}, {Kay}, {Kobayashi},
  {Kuang}, {Li}, {Lynch}, {L{\"u}ftinger}, {Luhmann}, {Maehara}, {Mlynczak},
  {Notsu}, {Osten}, {Ramirez}, {Rugheimer}, {Scheucher}, {Schlieder},
  {Shibata}, {Sousa-Silva}, {Stamenkovi{\'c}}, {Strangeway}, {Usmanov},
  {Vergados}, {Verkhoglyadova}, {Vidotto}, {Voytek}, {Way}, {Zank}, \&
  {Yamashiki}}]{2020IJAsB..19..136A}
{Airapetian}, V.~S., {Barnes}, R., {Cohen}, O., {et~al.} 2020, International
  Journal of Astrobiology, 19, 136, \dodoi{10.1017/S1473550419000132}

\bibitem[{{Allred} {et~al.}(2006){Allred}, {Hawley}, {Abbett}, \&
  {Carlsson}}]{2006ApJ...644..484A}
{Allred}, J.~C., {Hawley}, S.~L., {Abbett}, W.~P., \& {Carlsson}, M. 2006,
  \apj, 644, 484, \dodoi{10.1086/503314}

\bibitem[{{Alvarado-G{\'o}mez} {et~al.}(2018){Alvarado-G{\'o}mez}, {Drake},
  {Cohen}, {Moschou}, \& {Garraffo}}]{2018ApJ...862...93A}
{Alvarado-G{\'o}mez}, J.~D., {Drake}, J.~J., {Cohen}, O., {Moschou}, S.~P., \&
  {Garraffo}, C. 2018, \apj, 862, 93, \dodoi{10.3847/1538-4357/aacb7f}

\bibitem[{{Asai} {et~al.}(2012){Asai}, {Ichimoto}, {Kita}, {Kurokawa}, \&
  {Shibata}}]{2012PASJ...64...20A}
{Asai}, A., {Ichimoto}, K., {Kita}, R., {Kurokawa}, H., \& {Shibata}, K. 2012,
  \pasj, 64, 20, \dodoi{10.1093/pasj/64.1.20}

\bibitem[{{Bruzek}(1964)}]{1964ApJ...140..746B}
{Bruzek}, A. 1964, \apj, 140, 746, \dodoi{10.1086/147969}

\bibitem[{{Candelaresi} {et~al.}(2014){Candelaresi}, {Hillier}, {Maehara},
  {Brandenburg}, \& {Shibata}}]{2014ApJ...792...67C}
{Candelaresi}, S., {Hillier}, A., {Maehara}, H., {Brandenburg}, A., \&
  {Shibata}, K. 2014, \apj, 792, 67, \dodoi{10.1088/0004-637X/792/1/67}

\bibitem[{{Canfield} {et~al.}(1990){Canfield}, {Penn}, {Wulser}, \&
  {Kiplinger}}]{1990ApJ...363..318C}
{Canfield}, R.~C., {Penn}, M.~J., {Wulser}, J.-P., \& {Kiplinger}, A.~L. 1990,
  \apj, 363, 318, \dodoi{10.1086/169345}

\bibitem[{{Cargill} \& {Bradshaw}(2013)}]{2013ApJ...772...40C}
{Cargill}, P.~J., \& {Bradshaw}, S.~J. 2013, \apj, 772, 40,
  \dodoi{10.1088/0004-637X/772/1/40}

\bibitem[{{Chang} {et~al.}(2015){Chang}, {Byun}, \&
  {Hartman}}]{2015ApJ...814...35C}
{Chang}, S.~W., {Byun}, Y.~I., \& {Hartman}, J.~D. 2015, \apj, 814, 35,
  \dodoi{10.1088/0004-637X/814/1/35}

\bibitem[{{Claes} \& {Keppens}(2019)}]{2019A&A...624A..96C}
{Claes}, N., \& {Keppens}, R. 2019, \aap, 624, A96,
  \dodoi{10.1051/0004-6361/201834699}

\bibitem[{{Claes} {et~al.}(2020){Claes}, {Keppens}, \&
  {Xia}}]{2020A&A...636A.112C}
{Claes}, N., {Keppens}, R., \& {Xia}, C. 2020, \aap, 636, A112,
  \dodoi{10.1051/0004-6361/202037616}

\bibitem[{{Davenport}(2016)}]{2016ApJ...829...23D}
{Davenport}, J. R.~A. 2016, \apj, 829, 23, \dodoi{10.3847/0004-637X/829/1/23}

\bibitem[{{Dennis} \& {Zarro}(1993)}]{1993SoPh..146..177D}
{Dennis}, B.~R., \& {Zarro}, D.~M. 1993, \solphys, 146, 177,
  \dodoi{10.1007/BF00662178}

\bibitem[{{Drake} {et~al.}(2013){Drake}, {Cohen}, {Yashiro}, \&
  {Gopalswamy}}]{2013ApJ...764..170D}
{Drake}, J.~J., {Cohen}, O., {Yashiro}, S., \& {Gopalswamy}, N. 2013, \apj,
  764, 170, \dodoi{10.1088/0004-637X/764/2/170}

\bibitem[{{Emslie} {et~al.}(2012){Emslie}, {Dennis}, {Shih}, {Chamberlin},
  {Mewaldt}, {Moore}, {Share}, {Vourlidas}, \& {Welsch}}]{2012ApJ...759...71E}
{Emslie}, A.~G., {Dennis}, B.~R., {Shih}, A.~Y., {et~al.} 2012, \apj, 759, 71,
  \dodoi{10.1088/0004-637X/759/1/71}

\bibitem[{{Fisher} {et~al.}(1985){Fisher}, {Canfield}, \&
  {McClymont}}]{1985ApJ...289..414F}
{Fisher}, G.~H., {Canfield}, R.~C., \& {McClymont}, A.~N. 1985, \apj, 289, 414,
  \dodoi{10.1086/162901}

\bibitem[{{Fuhrmeister} {et~al.}(2018){Fuhrmeister}, {Czesla}, {Schmitt},
  {Jeffers}, {Caballero}, {Zechmeister}, {Reiners}, {Ribas}, {Amado},
  {Quirrenbach}, {B{\'e}jar}, {Galad{\'\i}-Enr{\'\i}quez}, {Guenther},
  {K{\"u}rster}, {Montes}, \& {Seifert}}]{2018A&A...615A..14F}
{Fuhrmeister}, B., {Czesla}, S., {Schmitt}, J.~H.~M.~M., {et~al.} 2018, \aap,
  615, A14, \dodoi{10.1051/0004-6361/201732204}

\bibitem[{{Gaidos} \& {Mann}(2014)}]{2014ApJ...791...54G}
{Gaidos}, E., \& {Mann}, A.~W. 2014, \apj, 791, 54,
  \dodoi{10.1088/0004-637X/791/1/54}

\bibitem[{{Hanaoka}(2003)}]{2003ApJ...596.1347H}
{Hanaoka}, Y. 2003, \apj, 596, 1347, \dodoi{10.1086/378123}

\bibitem[{{Harra-Murnion} {et~al.}(1998){Harra-Murnion}, {Schmieder}, {van
  Driel-Gesztelyi}, {Sato}, {Plunkett}, {Rudawy}, {Rompolt}, {Akioka}, {Sakao},
  \& {Ichimoto}}]{1998A&A...337..911H}
{Harra-Murnion}, L.~K., {Schmieder}, B., {van Driel-Gesztelyi}, L., {et~al.}
  1998, \aap, 337, 911

\bibitem[{{Hawley} {et~al.}(2014){Hawley}, {Davenport}, {Kowalski},
  {Wisniewski}, {Hebb}, {Deitrick}, \& {Hilton}}]{2014ApJ...797..121H}
{Hawley}, S.~L., {Davenport}, J. R.~A., {Kowalski}, A.~F., {et~al.} 2014, \apj,
  797, 121, \dodoi{10.1088/0004-637X/797/2/121}

\bibitem[{{Hawley} \& {Fisher}(1992)}]{1992ApJS...78..565H}
{Hawley}, S.~L., \& {Fisher}, G.~H. 1992, \apjs, 78, 565,
  \dodoi{10.1086/191640}

\bibitem[{{Houdebine} {et~al.}(1993){Houdebine}, {Foing}, {Doyle}, \&
  {Rodono}}]{1993A&A...274..245H}
{Houdebine}, E.~R., {Foing}, B.~H., {Doyle}, J.~G., \& {Rodono}, M. 1993, \aap,
  274, 245

\bibitem[{{Houdebine} {et~al.}(2016){Houdebine}, {Mullan}, {Paletou}, \&
  {Gebran}}]{2016ApJ...822...97H}
{Houdebine}, E.~R., {Mullan}, D.~J., {Paletou}, F., \& {Gebran}, M. 2016, \apj,
  822, 97, \dodoi{10.3847/0004-637X/822/2/97}

\bibitem[{{Howard} {et~al.}(2020){Howard}, {Corbett}, {Law}, {Ratzloff},
  {Galliher}, {Glazier}, {Gonzalez}, {Vasquez Soto}, {Fors}, {del Ser}, \&
  {Haislip}}]{2020ApJ...902..115H}
{Howard}, W.~S., {Corbett}, H., {Law}, N.~M., {et~al.} 2020, \apj, 902, 115,
  \dodoi{10.3847/1538-4357/abb5b4}

\bibitem[{{Hudson} {et~al.}(2006){Hudson}, {Wolfson}, \&
  {Metcalf}}]{2006SoPh..234...79H}
{Hudson}, H.~S., {Wolfson}, C.~J., \& {Metcalf}, T.~R. 2006, \solphys, 234, 79,
  \dodoi{10.1007/s11207-006-0056-y}

\bibitem[{{Ichimoto} \& {Kurokawa}(1984)}]{1984SoPh...93..105I}
{Ichimoto}, K., \& {Kurokawa}, H. 1984, \solphys, 93, 105,
  \dodoi{10.1007/BF00156656}

\bibitem[{{Janssens} \& {White}(1970)}]{1970SoPh...11..299J}
{Janssens}, T.~J., \& {White}, K.~P., I. 1970, \solphys, 11, 299,
  \dodoi{10.1007/BF00155229}

\bibitem[{{Johns-Krull} {et~al.}(1997){Johns-Krull}, {Hawley}, {Basri}, \&
  {Valenti}}]{1997ApJS..112..221J}
{Johns-Krull}, C.~M., {Hawley}, S.~L., {Basri}, G., \& {Valenti}, J.~A. 1997,
  \apjs, 112, 221, \dodoi{10.1086/313030}

\bibitem[{{Kane} {et~al.}(1983){Kane}, {Kai}, {Kosugi}, {Enome}, {Landecker},
  \& {McKenzie}}]{1983ApJ...271..376K}
{Kane}, S.~R., {Kai}, K., {Kosugi}, T., {et~al.} 1983, \apj, 271, 376,
  \dodoi{10.1086/161203}

\bibitem[{{Kawate} {et~al.}(2016){Kawate}, {Ishii}, {Nakatani}, {Ichimoto},
  {Asai}, {Morita}, \& {Masuda}}]{2016ApJ...833...50K}
{Kawate}, T., {Ishii}, T.~T., {Nakatani}, Y., {et~al.} 2016, \apj, 833, 50,
  \dodoi{10.3847/1538-4357/833/1/50}

\bibitem[{{Koller} {et~al.}(2021){Koller}, {Leitzinger}, {Temmer}, {Odert},
  {Beck}, \& {Veronig}}]{2021A&A...646A..34K}
{Koller}, F., {Leitzinger}, M., {Temmer}, M., {et~al.} 2021, \aap, 646, A34,
  \dodoi{10.1051/0004-6361/202039003}

\bibitem[{{Kowalski} {et~al.}(2022{\natexlab{a}}){Kowalski}, {Allred},
  {Carlsson}, {Kerr}, {Tremblay}, {Namekata}, {Kuridze}, \&
  {Uitenbroek}}]{2022arXiv220113349K}
{Kowalski}, A.~F., {Allred}, J.~C., {Carlsson}, M., {et~al.}
  2022{\natexlab{a}}, arXiv e-prints, arXiv:2201.13349.
\newblock \doarXiv{2201.13349}

\bibitem[{{Kowalski} {et~al.}(2022{\natexlab{b}}){Kowalski}, {Allred},
  {Carlsson}, {Kerr}, {Tremblay}, {Namekata}, {Kuridze}, \&
  {Uitenbroek}}]{2022ApJ...928..190K}
---. 2022{\natexlab{b}}, \apj, 928, 190, \dodoi{10.3847/1538-4357/ac5174}

\bibitem[{{Kowalski} {et~al.}(2017{\natexlab{a}}){Kowalski}, {Allred}, {Daw},
  {Cauzzi}, \& {Carlsson}}]{2017ApJ...836...12K}
{Kowalski}, A.~F., {Allred}, J.~C., {Daw}, A., {Cauzzi}, G., \& {Carlsson}, M.
  2017{\natexlab{a}}, \apj, 836, 12, \dodoi{10.3847/1538-4357/836/1/12}

\bibitem[{{Kowalski} {et~al.}(2015){Kowalski}, {Hawley}, {Carlsson}, {Allred},
  {Uitenbroek}, {Osten}, \& {Holman}}]{2015SoPh..290.3487K}
{Kowalski}, A.~F., {Hawley}, S.~L., {Carlsson}, M., {et~al.} 2015, \solphys,
  290, 3487, \dodoi{10.1007/s11207-015-0708-x}

\bibitem[{{Kowalski} {et~al.}(2009){Kowalski}, {Hawley}, {Hilton}, {Becker},
  {West}, {Bochanski}, \& {Sesar}}]{2009AJ....138..633K}
{Kowalski}, A.~F., {Hawley}, S.~L., {Hilton}, E.~J., {et~al.} 2009, \aj, 138,
  633, \dodoi{10.1088/0004-6256/138/2/633}

\bibitem[{{Kowalski} {et~al.}(2010){Kowalski}, {Hawley}, {Holtzman},
  {Wisniewski}, \& {Hilton}}]{2010ApJ...714L..98K}
{Kowalski}, A.~F., {Hawley}, S.~L., {Holtzman}, J.~A., {Wisniewski}, J.~P., \&
  {Hilton}, E.~J. 2010, \apjl, 714, L98, \dodoi{10.1088/2041-8205/714/1/L98}

\bibitem[{{Kowalski} {et~al.}(2013){Kowalski}, {Hawley}, {Wisniewski}, {Osten},
  {Hilton}, {Holtzman}, {Schmidt}, \& {Davenport}}]{2013ApJS..207...15K}
{Kowalski}, A.~F., {Hawley}, S.~L., {Wisniewski}, J.~P., {et~al.} 2013, \apjs,
  207, 15, \dodoi{10.1088/0067-0049/207/1/15}

\bibitem[{{Kowalski} {et~al.}(2016){Kowalski}, {Mathioudakis}, {Hawley},
  {Wisniewski}, {Dhillon}, {Marsh}, {Hilton}, \& {Brown}}]{2016ApJ...820...95K}
{Kowalski}, A.~F., {Mathioudakis}, M., {Hawley}, S.~L., {et~al.} 2016, \apj,
  820, 95, \dodoi{10.3847/0004-637X/820/2/95}

\bibitem[{{Kowalski} {et~al.}(2017{\natexlab{b}}){Kowalski}, {Allred},
  {Uitenbroek}, {Tremblay}, {Brown}, {Carlsson}, {Osten}, {Wisniewski}, \&
  {Hawley}}]{2017ApJ...837..125K}
{Kowalski}, A.~F., {Allred}, J.~C., {Uitenbroek}, H., {et~al.}
  2017{\natexlab{b}}, \apj, 837, 125, \dodoi{10.3847/1538-4357/aa603e}

\bibitem[{{Kretzschmar}(2011)}]{2011A&A...530A..84K}
{Kretzschmar}, M. 2011, \aap, 530, A84, \dodoi{10.1051/0004-6361/201015930}

\bibitem[{{Krucker} {et~al.}(2011){Krucker}, {Hudson}, {Jeffrey}, {Battaglia},
  {Kontar}, {Benz}, {Csillaghy}, \& {Lin}}]{2011ApJ...739...96K}
{Krucker}, S., {Hudson}, H.~S., {Jeffrey}, N.~L.~S., {et~al.} 2011, \apj, 739,
  96, \dodoi{10.1088/0004-637X/739/2/96}

\bibitem[{{Kurita} {et~al.}(2020){Kurita}, {Kino}, {Iwamuro}, {Ohta}, {Nogami},
  {Izumiura}, {Yoshida}, {Matsubayashi}, {Kuroda}, {Nakatani}, {Yamamoto},
  {Tsutsui}, {Iribe}, {Jikuya}, {Ohtani}, {Shibata}, {Takahashi}, {Tokoro},
  {Maihara}, \& {Nagata}}]{2020PASJ...72...48K}
{Kurita}, M., {Kino}, M., {Iwamuro}, F., {et~al.} 2020, \pasj, 72, 48,
  \dodoi{10.1093/pasj/psaa036}

\bibitem[{{Lacy} {et~al.}(1976){Lacy}, {Moffett}, \&
  {Evans}}]{1976ApJS...30...85L}
{Lacy}, C.~H., {Moffett}, T.~J., \& {Evans}, D.~S. 1976, \apjs, 30, 85,
  \dodoi{10.1086/190358}

\bibitem[{{Livshits} {et~al.}(1981){Livshits}, {Badalian}, {Kosovichev}, \&
  {Katsova}}]{1981SoPh...73..269L}
{Livshits}, M.~A., {Badalian}, O.~G., {Kosovichev}, A.~G., \& {Katsova}, M.~M.
  1981, \solphys, 73, 269, \dodoi{10.1007/BF00151682}

\bibitem[{{Longcope}(2014)}]{2014ApJ...795...10L}
{Longcope}, D.~W. 2014, \apj, 795, 10, \dodoi{10.1088/0004-637X/795/1/10}

\bibitem[{{Machado} {et~al.}(1989){Machado}, {Emslie}, \&
  {Avrett}}]{1989SoPh..124..303M}
{Machado}, M.~E., {Emslie}, A.~G., \& {Avrett}, E.~H. 1989, \solphys, 124, 303,
  \dodoi{10.1007/BF00156272}

\bibitem[{{Maehara} {et~al.}(2012){Maehara}, {Shibayama}, {Notsu}, {Notsu},
  {Nagao}, {Kusaba}, {Honda}, {Nogami}, \& {Shibata}}]{2012Natur.485..478M}
{Maehara}, H., {Shibayama}, T., {Notsu}, S., {et~al.} 2012, \nat, 485, 478,
  \dodoi{10.1038/nature11063}

\bibitem[{{Maehara} {et~al.}(2021){Maehara}, {Notsu}, {Namekata}, {Honda},
  {Kowalski}, {Katoh}, {Ohshima}, {Iida}, {Oeda}, {Murata}, {Yamanaka},
  {Takagi}, {Sasada}, {Akitaya}, {Ikuta}, {Okamoto}, {Nogami}, \&
  {Shibata}}]{2021PASJ...73...44M}
{Maehara}, H., {Notsu}, Y., {Namekata}, K., {et~al.} 2021, \pasj, 73, 44,
  \dodoi{10.1093/pasj/psaa098}

\bibitem[{{Matsubayashi} {et~al.}(2019){Matsubayashi}, {Ohta}, {Iwamuro},
  {Iwata}, {Kambe}, {Tsutsui}, {Izumiura}, {Yoshida}, \&
  {Hattori}}]{2019PASJ...71..102M}
{Matsubayashi}, K., {Ohta}, K., {Iwamuro}, F., {et~al.} 2019, \pasj, 71, 102,
  \dodoi{10.1093/pasj/psz087}

\bibitem[{{Morin} {et~al.}(2008){Morin}, {Donati}, {Petit}, {Delfosse},
  {Forveille}, {Albert}, {Auri{\`e}re}, {Cabanac}, {Dintrans}, {Fares},
  {Gastine}, {Jardine}, {Ligni{\`e}res}, {Paletou}, {Ramirez Velez}, \&
  {Th{\'e}ado}}]{2008MNRAS.390..567M}
{Morin}, J., {Donati}, J.~F., {Petit}, P., {et~al.} 2008, \mnras, 390, 567,
  \dodoi{10.1111/j.1365-2966.2008.13809.x}

\bibitem[{{Najita} \& {Orrall}(1970)}]{1970SoPh...15..176N}
{Najita}, K., \& {Orrall}, F.~Q. 1970, \solphys, 15, 176,
  \dodoi{10.1007/BF00149484}

\bibitem[{{Namekata} {et~al.}(2022{\natexlab{a}}){Namekata}, {Ichimoto},
  {Ishii}, \& {Shibata}}]{2022ApJ...933..209N}
{Namekata}, K., {Ichimoto}, K., {Ishii}, T.~T., \& {Shibata}, K.
  2022{\natexlab{a}}, \apj, 933, 209, \dodoi{10.3847/1538-4357/ac75cd}

\bibitem[{{Namekata} {et~al.}(2017){Namekata}, {Sakaue}, {Watanabe}, {Asai},
  {Maehara}, {Notsu}, {Notsu}, {Honda}, {Ishii}, {Ikuta}, {Nogami}, \&
  {Shibata}}]{2017ApJ...851...91N}
{Namekata}, K., {Sakaue}, T., {Watanabe}, K., {et~al.} 2017, \apj, 851, 91,
  \dodoi{10.3847/1538-4357/aa9b34}

\bibitem[{{Namekata} {et~al.}(2020){Namekata}, {Maehara}, {Sasaki}, {Kawai},
  {Notsu}, {Kowalski}, {Allred}, {Iwakiri}, {Tsuboi}, {Murata}, {Niwano},
  {Shiraishi}, {Adachi}, {Iida}, {Oeda}, {Honda}, {Tozuka}, {Katoh}, {Onozato},
  {Okamoto}, {Isogai}, {Kimura}, {Kojiguchi}, {Wakamatsu}, {Tampo}, {Nogami},
  \& {Shibata}}]{2020PASJ...72...68N}
{Namekata}, K., {Maehara}, H., {Sasaki}, R., {et~al.} 2020, \pasj, 72, 68,
  \dodoi{10.1093/pasj/psaa051}

\bibitem[{{Namekata} {et~al.}(2021){Namekata}, {Maehara}, {Honda}, {Notsu},
  {Okamoto}, {Takahashi}, {Takayama}, {Ohshima}, {Saito}, {Katoh}, {Tozuka},
  {Murata}, {Ogawa}, {Niwano}, {Adachi}, {Oeda}, {Shiraishi}, {Isogai}, {Seki},
  {Ishii}, {Ichimoto}, {Nogami}, \& {Shibata}}]{2022NatAs...6..241N}
{Namekata}, K., {Maehara}, H., {Honda}, S., {et~al.} 2021, Nature Astronomy, 6,
  241, \dodoi{10.1038/s41550-021-01532-8}

\bibitem[{{Namekata} {et~al.}(2022{\natexlab{b}}){Namekata}, {Maehara},
  {Honda}, {Notsu}, {Okamoto}, {Takahashi}, {Takayama}, {Ohshima}, {Saito},
  {Katoh}, {Tozuka}, {Murata}, {Ogawa}, {Niwano}, {Adachi}, {Oeda},
  {Shiraishi}, {Isogai}, {Nogami}, \& {Shibata}}]{2022ApJ...926L...5N}
---. 2022{\natexlab{b}}, \apjl, 926, L5, \dodoi{10.3847/2041-8213/ac4df0}

\bibitem[{{Neupert}(1968)}]{1968ApJ...153L..59N}
{Neupert}, W.~M. 1968, \apjl, 153, L59, \dodoi{10.1086/180220}

\bibitem[{{Newton} {et~al.}(2015){Newton}, {Charbonneau}, {Irwin}, \&
  {Mann}}]{2015ApJ...800...85N}
{Newton}, E.~R., {Charbonneau}, D., {Irwin}, J., \& {Mann}, A.~W. 2015, \apj,
  800, 85, \dodoi{10.1088/0004-637X/800/2/85}

\bibitem[{{Nizamov}(2019)}]{2019MNRAS.489.4338N}
{Nizamov}, B.~A. 2019, \mnras, 489, 4338, \dodoi{10.1093/mnras/stz2478}

\bibitem[{{Notsu} {et~al.}(2013){Notsu}, {Shibayama}, {Maehara}, {Notsu},
  {Nagao}, {Honda}, {Ishii}, {Nogami}, \& {Shibata}}]{2013ApJ...771..127N}
{Notsu}, Y., {Shibayama}, T., {Maehara}, H., {et~al.} 2013, \apj, 771, 127,
  \dodoi{10.1088/0004-637X/771/2/127}

\bibitem[{{Notsu} {et~al.}(2019){Notsu}, {Maehara}, {Honda}, {Hawley},
  {Davenport}, {Namekata}, {Notsu}, {Ikuta}, {Nogami}, \&
  {Shibata}}]{2019ApJ...876...58N}
{Notsu}, Y., {Maehara}, H., {Honda}, S., {et~al.} 2019, \apj, 876, 58,
  \dodoi{10.3847/1538-4357/ab14e6}

\bibitem[{{Okamoto} {et~al.}(2021){Okamoto}, {Notsu}, {Maehara}, {Namekata},
  {Honda}, {Ikuta}, {Nogami}, \& {Shibata}}]{2021ApJ...906...72O}
{Okamoto}, S., {Notsu}, Y., {Maehara}, H., {et~al.} 2021, \apj, 906, 72,
  \dodoi{10.3847/1538-4357/abc8f5}

\bibitem[{{Otsu} {et~al.}(2022){Otsu}, {Asai}, {Ichimoto}, {Ishii}, \&
  {Namekata}}]{2022ApJ...939...98O}
{Otsu}, T., {Asai}, A., {Ichimoto}, K., {Ishii}, T.~T., \& {Namekata}, K. 2022,
  \apj, 939, 98, \dodoi{10.3847/1538-4357/ac9730}

\bibitem[{{Priest} \& {Forbes}(2002)}]{2002A&ARv..10..313P}
{Priest}, E.~R., \& {Forbes}, T.~G. 2002, \aapr, 10, 313,
  \dodoi{10.1007/s001590100013}

\bibitem[{{Reep} {et~al.}(2020){Reep}, {Antolin}, \&
  {Bradshaw}}]{2020ApJ...890..100R}
{Reep}, J.~W., {Antolin}, P., \& {Bradshaw}, S.~J. 2020, \apj, 890, 100,
  \dodoi{10.3847/1538-4357/ab6bdc}

\bibitem[{{Ricker} {et~al.}(2015){Ricker}, {Winn}, {Vanderspek}, {Latham},
  {Bakos}, {Bean}, {Berta-Thompson}, {Brown}, {Buchhave}, {Butler}, {Butler},
  {Chaplin}, {Charbonneau}, {Christensen-Dalsgaard}, {Clampin}, {Deming},
  {Doty}, {De Lee}, {Dressing}, {Dunham}, {Endl}, {Fressin}, {Ge}, {Henning},
  {Holman}, {Howard}, {Ida}, {Jenkins}, {Jernigan}, {Johnson}, {Kaltenegger},
  {Kawai}, {Kjeldsen}, {Laughlin}, {Levine}, {Lin}, {Lissauer}, {MacQueen},
  {Marcy}, {McCullough}, {Morton}, {Narita}, {Paegert}, {Palle}, {Pepe},
  {Pepper}, {Quirrenbach}, {Rinehart}, {Sasselov}, {Sato}, {Seager},
  {Sozzetti}, {Stassun}, {Sullivan}, {Szentgyorgyi}, {Torres}, {Udry}, \&
  {Villasenor}}]{2015JATIS...1a4003R}
{Ricker}, G.~R., {Winn}, J.~N., {Vanderspek}, R., {et~al.} 2015, Journal of
  Astronomical Telescopes, Instruments, and Systems, 1, 014003,
  \dodoi{10.1117/1.JATIS.1.1.014003}

\bibitem[{{Ruan} {et~al.}(2021){Ruan}, {Zhou}, \&
  {Keppens}}]{2021ApJ...920L..15R}
{Ruan}, W., {Zhou}, Y., \& {Keppens}, R. 2021, \apjl, 920, L15,
  \dodoi{10.3847/2041-8213/ac27b0}

\bibitem[{{Shibata} \& {Magara}(2011)}]{2011LRSP....8....6S}
{Shibata}, K., \& {Magara}, T. 2011, Living Reviews in Solar Physics, 8, 6,
  \dodoi{10.12942/lrsp-2011-6}

\bibitem[{{Shibata} \& {Yokoyama}(2002)}]{2002ApJ...577..422S}
{Shibata}, K., \& {Yokoyama}, T. 2002, \apj, 577, 422, \dodoi{10.1086/342141}

\bibitem[{{Shibata} {et~al.}(2013){Shibata}, {Isobe}, {Hillier}, {Choudhuri},
  {Maehara}, {Ishii}, {Shibayama}, {Notsu}, {Notsu}, {Nagao}, {Honda}, \&
  {Nogami}}]{2013PASJ...65...49S}
{Shibata}, K., {Isobe}, H., {Hillier}, A., {et~al.} 2013, \pasj, 65, 49,
  \dodoi{10.1093/pasj/65.3.49}

\bibitem[{{Shibayama} {et~al.}(2013){Shibayama}, {Maehara}, {Notsu}, {Notsu},
  {Nagao}, {Honda}, {Ishii}, {Nogami}, \& {Shibata}}]{2013ApJS..209....5S}
{Shibayama}, T., {Maehara}, H., {Notsu}, S., {et~al.} 2013, \apjs, 209, 5,
  \dodoi{10.1088/0067-0049/209/1/5}

\bibitem[{{Somov} {et~al.}(1982){Somov}, {Sermulina}, \&
  {Spektor}}]{1982SoPh...81..281S}
{Somov}, B.~V., {Sermulina}, B.~J., \& {Spektor}, A.~R. 1982, \solphys, 81,
  281, \dodoi{10.1007/BF00151302}

\bibitem[{{Soubiran} {et~al.}(2018){Soubiran}, {Jasniewicz}, {Chemin},
  {Zurbach}, {Brouillet}, {Panuzzo}, {Sartoretti}, {Katz}, {Le Campion},
  {Marchal}, {Hestroffer}, {Th{\'e}venin}, {Crifo}, {Udry}, {Cropper},
  {Seabroke}, {Viala}, {Benson}, {Blomme}, {Jean-Antoine}, {Huckle}, {Smith},
  {Baker}, {Damerdji}, {Dolding}, {Fr{\'e}mat}, {Gosset}, {Guerrier}, {Guy},
  {Haigron}, {Jan{\ss}en}, {Plum}, {Fabre}, {Lasne}, {Pailler}, {Panem},
  {Riclet}, {Royer}, {Tauran}, {Zwitter}, {Gueguen}, \&
  {Turon}}]{2018A&A...616A...7S}
{Soubiran}, C., {Jasniewicz}, G., {Chemin}, L., {et~al.} 2018, \aap, 616, A7,
  \dodoi{10.1051/0004-6361/201832795}

\bibitem[{{Svestka}(1962)}]{1962BAICz..13..236S}
{Svestka}, Z. 1962, Bulletin of the Astronomical Institutes of Czechoslovakia,
  13, 236

\bibitem[{{Svestka}(1963)}]{1963BAICz..14..234S}
---. 1963, Bulletin of the Astronomical Institutes of Czechoslovakia, 14, 234

\bibitem[{{van Maanen}(1945)}]{1945PASP...57..216V}
{van Maanen}, A. 1945, \pasp, 57, 216, \dodoi{10.1086/125730}

\bibitem[{{Vida} {et~al.}(2019){Vida}, {Leitzinger}, {Kriskovics}, {Seli},
  {Odert}, {Kov{\'a}cs}, {Korhonen}, \& {van
  Driel-Gesztelyi}}]{2019A&A...623A..49V}
{Vida}, K., {Leitzinger}, M., {Kriskovics}, L., {et~al.} 2019, \aap, 623, A49,
  \dodoi{10.1051/0004-6361/201834264}

\bibitem[{{{\v{S}}vestka} {et~al.}(1962){{\v{S}}vestka}, {Kopeck{\'y}}, \&
  {Blaha}}]{1962BAICz..13...37S}
{{\v{S}}vestka}, Z., {Kopeck{\'y}}, M., \& {Blaha}, M. 1962, Bulletin of the
  Astronomical Institutes of Czechoslovakia, 13, 37

\bibitem[{{Wu} {et~al.}(2022){Wu}, {Chen}, {Tian}, {Zhang}, {Shi}, {He}, {Lu},
  {Xu}, \& {Wang}}]{2022arXiv220302292W}
{Wu}, Y., {Chen}, H., {Tian}, H., {et~al.} 2022, arXiv e-prints,
  arXiv:2203.02292.
\newblock \doarXiv{2203.02292}

\bibitem[{{Xu} {et~al.}(2006){Xu}, {Cao}, {Liu}, {Yang}, {Jing}, {Denker},
  {Emslie}, \& {Wang}}]{2006ApJ...641.1210X}
{Xu}, Y., {Cao}, W., {Liu}, C., {et~al.} 2006, \apj, 641, 1210,
  \dodoi{10.1086/500632}

\bibitem[{{Yamashiki} {et~al.}(2019){Yamashiki}, {Maehara}, {Airapetian},
  {Notsu}, {Sato}, {Notsu}, {Kuroki}, {Murashima}, {Sato}, {Namekata},
  {Sasaki}, {Scott}, {Bando}, {Nashimoto}, {Takagi}, {Ling}, {Nogami}, \&
  {Shibata}}]{2019ApJ...881..114Y}
{Yamashiki}, Y.~A., {Maehara}, H., {Airapetian}, V., {et~al.} 2019, \apj, 881,
  114, \dodoi{10.3847/1538-4357/ab2a71}

\end{thebibliography}
\bibliographystyle{aasjournal}

\end{document}